\documentclass[conference, letterpaper]{IEEEtran}
\IEEEoverridecommandlockouts    

\usepackage[T1]{fontenc}
\usepackage[final]{microtype}
\usepackage{indentfirst}

\usepackage{multirow}
\usepackage{amsmath}
\usepackage{amssymb}
\usepackage{graphicx}
\graphicspath{{./Figures/}}

\usepackage{array}
\usepackage{tabularx}
\usepackage{booktabs}
\usepackage{makecell}
\usepackage[table]{xcolor}
\definecolor{OursRow}{HTML}{E1F2EB}
\definecolor{AblRow}{HTML}{F8E7D8}
\newcommand{\cellO}[1]{\cellcolor{OursRow}#1}
\newcommand{\cellAb}[2]{\cellcolor{AblRow!#1}#2}
\newcommand{\std}[1]{{\tiny\color{gray!90}$\pm$#1}}
\newcolumntype{L}{>{\raggedright\arraybackslash}X}

\usepackage{url}
\usepackage{xurl}

\usepackage[letterpaper, left=54pt, right=54pt, top=54pt, bottom=54pt]{geometry}

\usepackage{float}  

\usepackage[caption=false,font=footnotesize]{subfig}

\usepackage{tikz}
\usepackage{pgfplots}
\pgfplotsset{compat=1.18}

\usepackage{algorithm}
\usepackage{algpseudocode}

\providecommand{\IEEEkeywords}[1]{%
	\par\noindent\textbf{Index Terms---}#1\par%
}

\newcommand{\ThanksPar}{\par}

\title{\vspace{18pt}\LARGE \bf SpinFlow: A Physics-Informed Spin Field Framework for Traffic Phase Inference and Transition Detection}

\author{Haopeng Deng, Fucheng Zheng, and Xinhai Xia\thanks{H. Deng and F. Zheng contributed equally. X. Xia is the corresponding author. All authors are with the School of Future Transportation, Guangzhou Maritime University, Guangzhou, China (e-mails: qianyhp@gmail.com; zhengfucheng2002@163.com; xiaxinhai@gzmtu.edu.cn).\ThanksPar This work was supported by the Key Area Special Projects for General Higher Education Institutions in Guangdong Province (2022ZDZX1021, 2024ZDZX1033), and by the Research Project of the Bureau of Education of Guangzhou Municipality (2024312023).}}

\begin{document}
	\maketitle
	\begin{tikzpicture}[remember picture,overlay]
		\node[anchor=north west, align=left, font=\footnotesize\sffamily\bfseries]
		at ([xshift=54pt,yshift=-24pt]current page.north west)
		{2026 IEEE International Conference on Intelligent Transportation Systems (ITSC)\\
		September 15--18, 2026, Naples, Italy};
	\end{tikzpicture}
	\thispagestyle{empty}
	\pagestyle{empty}
	
	\begin{abstract}
		Active traffic management (ATM) is frequently hindered by traditional macroscopic models and rigid empirical thresholds that fail to capture metastable phase precursors, resulting in delayed, reactive interventions. To address this, we propose SpinFlow, a physics-informed spin-field framework unifying Kerner's three-phase theory with statistical physics for continuous macroscopic traffic phase inference. Inspired by the Heisenberg model, SpinFlow parametrizes spatially varying phase weights via a latent spin vector and a competitive-equilibrium mapping, allowing synchronized flow to emerge naturally. A physics-regularized Expectation-Maximization algorithm inverts this latent structure from high-resolution trajectories, jointly optimizing the spin field while softly enforcing mass conservation and spatial smoothness. We introduce the Phase Equilibrium Degree (PED) to quantify structural alignment and topologically localize phase-transition points. Across four real-world trajectory datasets, SpinFlow achieves $R_q^2$ up to 0.940, PED drops of 94.9--100\,\%, and interpretable phase maps that outperform three heterogeneous baselines on forward accuracy, physics consistency, and bottleneck localization. SpinFlow pinpoints congestion nucleation without prior network topology, yielding a data-driven, physics-consistent trigger for ATM.
	\end{abstract}
	
	\IEEEkeywords{Traffic phase inference; three-phase traffic theory; fundamental diagram; physics-informed learning.}
	\section{Introduction}
	\label{sec:introduction}

Proactive macroscopic traffic phase inference and transition identification are essential for enhancing operational efficiency and safety in Intelligent Transportation Systems (ITS). Classical macroscopic traffic flow theory originated from the fundamental diagram (FD) for traffic flow, which has evolved over nearly 90 years \cite{turner201175}---e.g., Greenshields' linear model and logarithmic or exponential variants---establishing the foundational density-flow-speed relationships. These inspired the Lighthill-Whitham-Richards (LWR) kinematic wave model \cite{lighthill1955kinematic, richards1956shock} that treats traffic as a compressible fluid obeying conservation laws. To ease the computational burden of the resulting PDEs, Daganzo proposed the Cell Transmission Model (CTM) \cite{daganzo1994cell}; second-order models such as Payne-Whitham (PW) \cite{payne1973} and Aw-Rascle-Zhang (AWZ) \cite{aw2000resurrection, zhang2002non} add velocity dynamics to capture non-equilibrium waves. Despite their maturity and utility, these two-phase (free vs. congested) frameworks with a single FD treat complex traffic dynamics isotropically. Consequently, they lack the endogenous mechanisms required to intrinsically explain key empirical phenomena---including hysteresis, capacity drop, multi-stability, and the spontaneous, nonlinear jumps between free and congested flow---owing to their strict reliance on continuous-medium and static-equilibrium assumptions.

To address these limitations, Kerner introduced three-phase traffic theory \cite{kerner2004physics, kerner2019statistical}, distinguishing free flow (F), synchronized flow (S), and wide moving jam (J). The theory characterizes the F$\to$S transition as a first-order phase change, with wide moving jams nucleating internally within synchronized flow; the transition zone exhibits pronounced metastability, stochastic triggering, and spatiotemporal competition \cite{wiering2022statistical, chen2024empirical, zhang2025traffic}. Empirical validations using trajectory data have extensively confirmed these features at geometric bottlenecks \cite{palmer2011reconstruction, xu2015safety}. Complementing these empirical advances, the statistical-physics perspective frames traffic as an emergent many-particle system \cite{helbing2001traffic, chowdhury2000statistical}. Nagatani's lattice hydrodynamic models \cite{nagatani2002physics, nagatani1999tdgl} revealed jamming transitions and critical phenomena. Notably, the structural isomorphism between traffic phase competition and lattice spin systems provides a particularly powerful analytical lens. While the Ising model \cite{ising1925beitrag} captures binary discrete states, the Heisenberg model \cite{heisenberg1932bau} extends this to continuous three-dimensional vector representations, effectively overcoming the theoretical bottlenecks of hard-boundary classifications. Together with Boltzmann statistics \cite{boltzmann1872weitere} for phase probabilities and Landau theory \cite{landau1937theory} for order-parameter stability boundaries, these physics-inspired paradigms offer a rigorous mathematical foundation. The Expectation-Maximization (EM) algorithm \cite{dempster1977maximum} and its variational free-energy interpretation \cite{neal1998em} provide the computational engine for such latent-variable inversion. While recent trajectory-driven calibrations \cite{he2025constructing} and hybrid physics-informed machine learning (PIML) \cite{zhang2023review, pan2023ising} have improved modeling robustness under sparse data, their reliance on neural backbones often sacrifices the strict interpretability inherent to fully transparent white-box models.

Despite these substantial advances, two critical gaps, persisting over 20 years since Kerner, severely limit engineering deployment and ATM. First, a persistent disconnection remains between microscopic phase-transition mechanisms and macroscopic, computable states. Existing macroscopic models and three-phase simulations excel at post-hoc reproduction but rarely translate complex latent physics into interpretable, spatially continuous phase representations. Second, the quantitative identification of critical equilibrium states remains elusive. Conventional state-recognition methodologies predominantly rely on hard empirical thresholds (e.g., rigid velocity or density cutoffs), which inherently fail to capture the subtle pre-transition precursors, structural entropy, and metastable fluctuations characterizing mixed-phase regimes. Consequently, modern control systems often intervene only after a shockwave has fully formed, entirely missing the narrow, low-cost window for preventive interventions.

This paper presents SpinFlow, a physics-informed spin-field framework that unifies Kerner's three-phase theory with statistical-physics latent-variable modeling for continuous macroscopic traffic phase inference and transition identification. Inspired by the Heisenberg model's continuous symmetry, SpinFlow parametrizes spatially varying phase weights $\boldsymbol{\pi}(x)$ via a latent spin vector $\mathbf{s}(x) \in \mathbb{R}^3$ and a competitive-equilibrium mapping. Free-jam competition bias and strength drive this mapping, allowing synchronized flow to emerge naturally as the balanced equilibrium outcome. A physics-regularized EM algorithm inverts the latent structure from high-resolution trajectory data \cite{feng2025ubiquitous, krajewski2018highd, coifman2017ngsim}. Finally, the framework introduces the Phase Equilibrium Degree (PED) to quantify structural alignment and to topologically localize quasi-equilibrium regions and phase-transition points, providing a data-driven, physics-consistent understanding of congestion nucleation and a computable trigger for ATM.

Experiments show tight in-sample forward consistency and interpretable phase maps; detected transitions are consistent with documented bottleneck features. A companion open-source repository is publicly maintained at \url{https://github.com/QianyHP/spinflow-traffic} under the MIT License, including evaluation scripts, visualization utilities, and extended analyses to support reproducibility. The remainder of the paper is organized as follows: Sec.~II details observations and phase prototypes; Sec.~III presents the SpinFlow model; Sec.~IV describes the EM inference and PED-based detection; Sec.~V reports experimental validation; and Sec.~VI concludes with future directions.
	
\section{Observations and Phase Prototypes}
\label{sec:observation_prototypes}

\subsection{Trajectory Representation and Edie Macroscopic Variables}
Vehicle trajectories $\mathcal{Z}$ collected on a road segment $x \in [0,L]$ over $t \in [t_0, t_0+\Delta T]$ form the basis of our analysis. 
For the $n$th trajectory $x_n(t)$ with speed $u_n(t)=\dot{x}_n(t)$ and any spatiotemporal region $\mathcal{A} \subset \mathbb{R}^2$, the total traveled distance and total time spent are
\begin{equation}
	d(\mathcal{A})=\sum_{n} d_n(\mathcal{A}), \qquad \tau(\mathcal{A})=\sum_{n}\tau_n(\mathcal{A}),
	\label{eq:edie_dtau}
\end{equation}
where $d_n(\mathcal{A})$ and $\tau_n(\mathcal{A})$ denote the spatial and temporal projection lengths of trajectory $n$ inside $\mathcal{A}$. 
Edie's macroscopic variables \cite{edie1963} follow as
\begin{equation}
	\rho(\mathcal{A})=\frac{\tau(\mathcal{A})}{|\mathcal{A}|}, \qquad q(\mathcal{A})=\frac{d(\mathcal{A})}{|\mathcal{A}|}, \qquad v(\mathcal{A})=\frac{q(\mathcal{A})}{\rho(\mathcal{A})},
	\label{eq:edie_rqv}
\end{equation}
with $|\mathcal{A}|$ representing the area of $\mathcal{A}$; reported values use $\rho$ in veh/km, $q$ in veh/h, and $v$ in km/h.
Since large rectangular windows obscure transient waves, eq. \eqref{eq:edie_rqv} is evaluated on sampled local regions that approximate quasi-stationarity.

\subsection{Quasi-Stationary Sampling via Parallelograms}
Building on the approach of He and Wu \cite{he2025constructing}, sampling employs parallelograms $\{\mathcal{P}_p\}_{p=1}^{N}$ in the $(t,x)$ plane to capture local homogeneity. 
To capture wave propagation, the geometry of $\mathcal{P}_p$ is dynamically adapted: long edges align with the congestion wave speed while short edges follow the target speed $v^{\ast}$, ensuring trajectories inside $\mathcal{P}_p$ are nearly parallel.

\begin{figure}[h]
\centering
\includegraphics[width=\columnwidth]{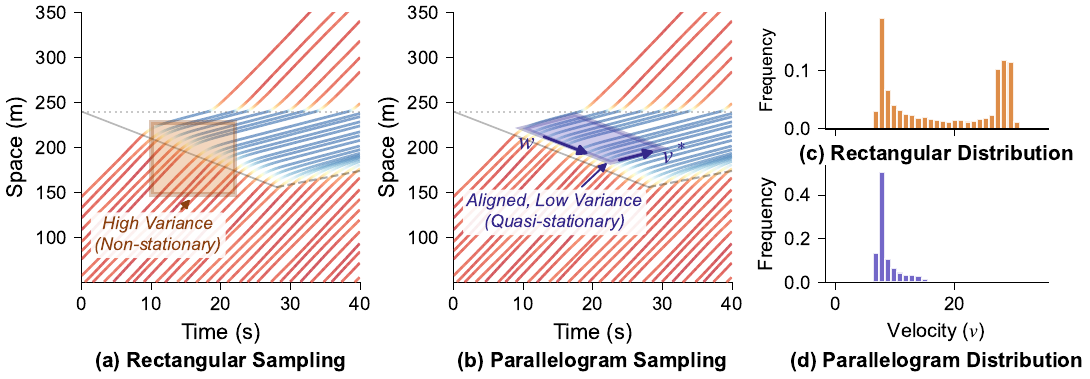}
\caption{(a)~Rectangular window straddles the shockwave: bimodal velocity distribution (high~$\sigma^2$). (b)~Parallelogram aligned with wave speed~$w$ and vehicle speed~$v^{\ast}$: quasi-stationary region (low~$\sigma^2$). (c,\,d)~Velocity histograms.}
\label{fig:sampling_mechanism}
\end{figure}

Computation for each candidate $\mathcal{P}_p$ yields
\begin{equation}
	(\rho_p,q_p,v_p)=\big(\rho(\mathcal{P}_p), q(\mathcal{P}_p), v(\mathcal{P}_p)\big),
	\label{eq:parallelogram_rqv}
\end{equation}
recorded at the spatial center $x_p$. 
This process generates the observation vector
\begin{equation}
	\mathbf{y}_p=(\rho_p,q_p,v_p,x_p).
	\label{eq:obs_vector}
\end{equation}
A stability metric evaluates each region by penalizing internal speed fluctuations and deviations from $v^{\ast}$, given by
\begin{equation}
	\mathrm{Score}(\mathcal{P}_p) = w_{\mathrm{CV}} \cdot \mathrm{CV}(\mathcal{P}_p) + w_{\mathrm{NAE}} \cdot \mathrm{NAE}(\mathcal{P}_p),
	\label{eq:score}
\end{equation}
where $\mathrm{CV}(\mathcal{P}_p)$ is the coefficient of variation of speeds and $\mathrm{NAE}(\mathcal{P}_p)$ is the normalized absolute error relative to $v^{\ast}$. 
An Otsu threshold $\eta$ \cite{otsu1979}, derived from the empirical score distribution with interquartile range $\mathrm{IQR}$, maps scores to continuous quality weights via
\begin{equation}
	w_p = \frac{1}{1 + \exp\big(\beta_w(\mathrm{Score}(\mathcal{P}_p)-\eta)\big)}, \qquad \beta_w=\ln(9)/\mathrm{IQR}.
	\label{eq:soft_weight}
\end{equation}
This logistic weighting suppresses low-quality regions ($w_p \approx 0$) during prototype fitting and inference.
Fig.~\ref{fig:sampling_mechanism} contrasts the parallelogram sampling with traditional rectangular sampling, showing the reduction in intra-window velocity variance.

\subsection{Three-Phase Prototype Fundamental Diagrams}
Three phase prototypes $\mathcal{G}=\{F,S,J\}$ represent free flow (F), synchronized flow (S), and wide moving jam (J), respectively. 
Each phase $g \in \mathcal{G}$ follows a triangular FD in the sense of Newell's kinematic wave theory \cite{newell1993}:
\begin{equation}
	\begin{aligned}
		q_g(\rho)=\min\Big\{v_f^{(g)}\rho,\; Q^{(g)},\; |w^{(g)}|\big(\rho_{\mathrm{jam}}^{(g)}-\rho\big)\Big\},\\
		\rho \in [0,\rho_{\mathrm{jam}}^{(g)}].
	\end{aligned}
	\label{eq:triangular_fd}
\end{equation}
The parameter vector $\boldsymbol{\theta}_g=(v_f^{(g)}, w^{(g)}, \rho_{\mathrm{jam}}^{(g)}, Q^{(g)})$, where $Q^{(g)}$ is capacity, respects the constraints $v_f^{(g)}>0$, $w^{(g)}<0$, $\rho_{\mathrm{jam}}^{(g)}>0$, and $Q^{(g)}>0$. 
Phase attractors, identified by K-Means clustering in the normalized $(\rho,v)$ space, partition observations into clusters $\{\mathcal{D}_g\}_{g\in\mathcal{G}}$. 
Weighted least squares then fits each prototype,
\begin{equation}
	\boldsymbol{\theta}_g^{\ast}=\arg\min_{\boldsymbol{\theta}_g}\sum_{p \in \mathcal{D}_g}\omega_p\Big(q_g(\rho_p;\boldsymbol{\theta}_g)-q_p\Big)^2,
	\label{eq:prototype_fit}
\end{equation}
where $\omega_p=w_p$ denotes the quasi-stationary weight from \eqref{eq:soft_weight}. 
These fitted prototypes form the phase-conditioned basis for the spatial mixture model.

\section{SpinFlow Model: Latent Spin Field and Competitive-Equilibrium Mapping}
\label{sec:spin_mapping}

\subsection{Spatial Mixture FD and Latent Spin Field}
Spatial heterogeneity is modeled by mixing the three phase prototypes with position-dependent weights $\boldsymbol{\pi}(x)=(\pi_F(x),\pi_S(x),\pi_J(x))$. 
The spatial mixture FD for density $\rho$ at location $x$ becomes
\begin{equation}
	q(\rho;x)=\sum_{g \in \mathcal{G}} \pi_g(x)\, q_g(\rho),
	\label{eq:mixture_fd}
\end{equation}
subject to $\pi_g(x)\ge 0$ and $\sum_{g\in\mathcal{G}}\pi_g(x)=1$.
Speed from the mixture follows implicitly as
\begin{equation}
	v(\rho;x)=
	\begin{cases}
		\dfrac{q(\rho;x)}{\rho}, & \rho>0,\\[2pt]
		\sum_{g\in\mathcal{G}}\pi_g(x)\, v_f^{(g)}, & \rho=0.
	\end{cases}
	\label{eq:mixture_speed}
\end{equation}
A latent spin field $\mathbf{s}(x)=(s_x(x),s_y(x),s_z(x)) \in \mathbb{R}^3$ parametrizes $\boldsymbol{\pi}(x)$. 
This spin vector offers a continuous representation of phase attractors while permitting mixed states in transition regions. 
Crucially, the magnitude $|\mathbf{s}(x)|$ remains unconstrained; the inversion process increases $|\mathbf{s}(x)|$ where data support a deterministic phase and keeps $|\mathbf{s}(x)|$ small where evidence is diffuse. We refer to this unconstrained spin magnitude as \emph{Phase Unfolding}.

\subsection{Competitive Mapping from Spin to Phase Weights}
We construct the mapping from $\mathbf{s}(x)$ to $\boldsymbol{\pi}(x)$ using three preference scores that explicitly encode the competition between free and jam phases. 
These scores are defined as
\begin{equation}
	\begin{aligned}
		h_F(\mathbf{s}(x)) &= s_z(x)-s_y(x),\\
		h_J(\mathbf{s}(x)) &= s_y(x)-s_z(x),\\
		h_S(\mathbf{s}(x)) &= s_x(x)-|s_z(x)-s_y(x)|.
	\end{aligned}
	\label{eq:preference_scores}
\end{equation}
The scores $h_F$ and $h_J$ form an antisymmetric pair ($h_J=-h_F$), driving the primary free--jam dichotomy. 
In contrast, $h_S$ incorporates a penalty term $|s_z-s_y|$, suppressing phase S when the free--jam competition is fierce.
A Boltzmann (Softmax) function then generates the weights
\begin{equation}
	\pi_g(x)=\frac{\exp\!\big(\beta\, h_g(\mathbf{s}(x))\big)}{\sum\limits_{g' \in \mathcal{G}} \exp\!\big(\beta\, h_{g'}(\mathbf{s}(x))\big)}, \qquad g\in\mathcal{G},
	\label{eq:softmax_mapping}
\end{equation}
where $\beta>0$ is the inverse temperature (distinct from $\beta_w$ in \eqref{eq:soft_weight}). 
This construction enforces the probability constraints on $\boldsymbol{\pi}(x)$ by design.

\subsection{Physical Interpretation: Bias and Competition Strength}
\begin{figure}[h]
\centering
\includegraphics[width=\columnwidth]{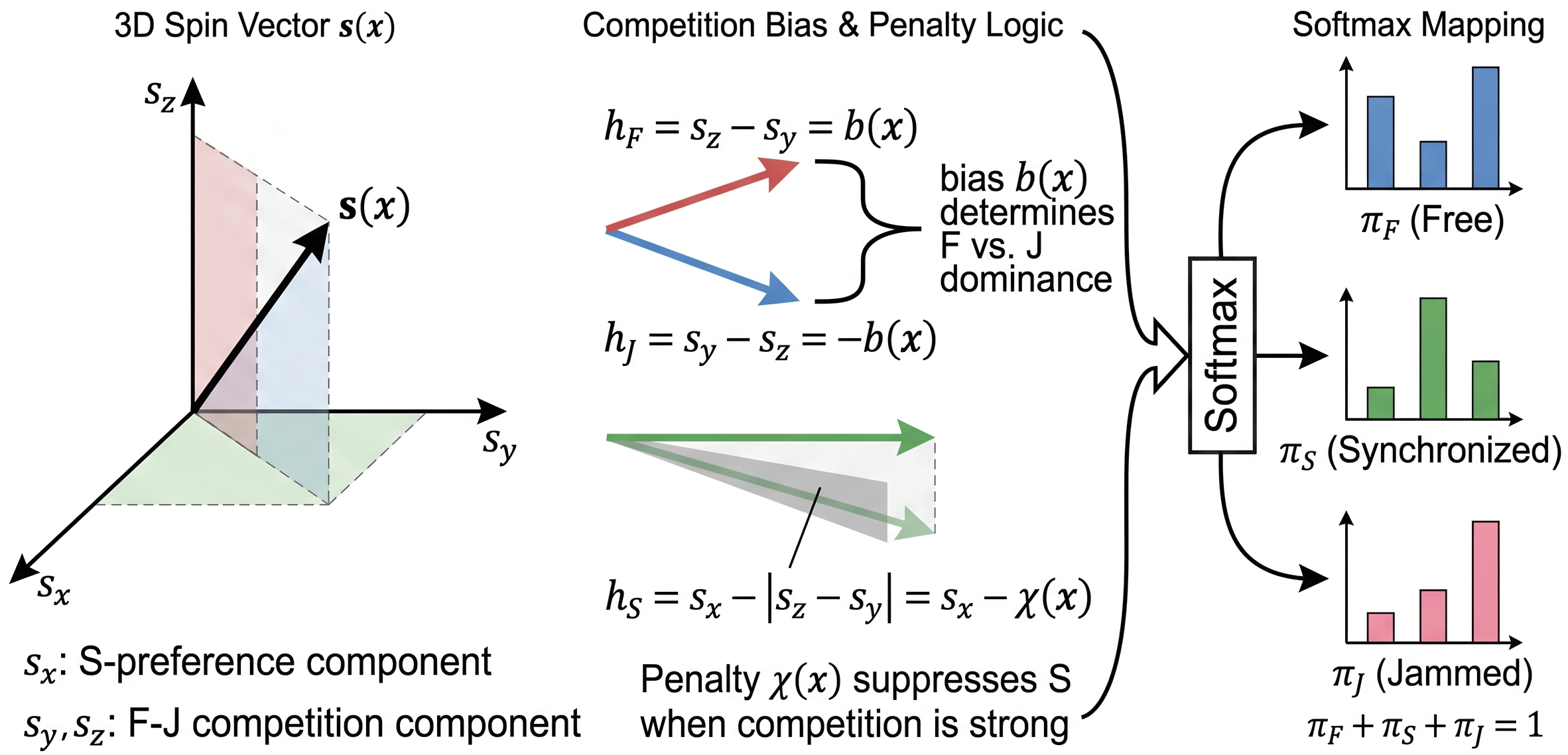}
\caption{Competitive-equilibrium mapping: bias $b(x)$ and penalty $\chi(x)$ govern preference scores $h_g$, which softmax maps to phase weights $\boldsymbol{\pi}(x)$.}
\label{fig:competition_mapping}
\end{figure}

The spin components admit a physical interpretation through the competition bias and competition strength,
\begin{equation}
	b(x)=s_z(x)-s_y(x), \qquad \chi(x)=|b(x)|.
	\label{eq:bias_strength}
\end{equation}
Substituting these into \eqref{eq:preference_scores} yields
\begin{equation}
	h_F=b, \qquad h_J=-b, \qquad h_S=s_x-\chi.
	\label{eq:score_rewrite}
\end{equation}
The sign of $b(x)$ dictates whether $F$ or $J$ dominates: $b(x)\gg 0$ implies $h_F\gg 0$ ($\pi_F(x)\approx 1$), whereas $b(x)\ll 0$ implies $h_J\gg 0$ ($\pi_J(x)\approx 1$). 
Phase~S is favored when free--jam competition is nearly balanced. As $b(x)\to 0$, $\chi(x)$ vanishes and $h_S \approx s_x$. 
If $s_x(x)>0$, $\pi_S(x)$ dominates and synchronized flow emerges from competitive equilibrium. Fig.~\ref{fig:competition_mapping} summarizes this spin-to-phase pipeline.

\section{Physics-Informed EM Inference and Transition Detection}
\label{sec:inference_identification}

\subsection{Generative Model and EM Responsibilities}
Each FD observation point $\mathbf{y}_p=(\rho_p,q_p,v_p,x_p)$ is modeled as a sample drawn from a mixture with latent phase label $z_p \in \mathcal{G}=\{F,S,J\}$. 
Conditional on $z_p=g$, the flow $q_p$ is assumed to follow a Gaussian distribution centered at the prototype $q_g(\rho_p;\boldsymbol{\theta}_g)$,
\begin{equation}
	q_p = q_g(\rho_p;\boldsymbol{\theta}_g) + \varepsilon_{q,p}, \qquad \varepsilon_{q,p}\sim\mathcal{N}(0,\sigma_q^2).
	\label{eq:flow_noise}
\end{equation}
Marginalizing over $z_p$ with the spatial mixture weights $\pi_g(x_p)$ yields the likelihood
\begin{equation}
	p(q_p\mid \rho_p,x_p)=\sum_{g\in\mathcal{G}} \pi_g(x_p)\,\mathcal{N}\!\big(q_p;\,q_g(\rho_p;\boldsymbol{\theta}_g),\,\sigma_q^2\big).
	\label{eq:mixture_likelihood}
\end{equation}
Neal and Hinton \cite{neal1998em} showed that EM minimizes a variational free energy, under which the E-step is the optimal posterior. The E-step at iteration $t$ computes the posterior responsibilities
\begin{equation}
	\begin{aligned}
		\gamma_{p,g}^{(t)}
		&=\Pr\!\big(z_p=g \mid q_p,\rho_p,x_p;\mathbf{s}^{(t)},\boldsymbol{\theta}^{(t)}\big)\\
		&=\frac{\pi_g^{(t)}(x_p)\,\mathcal{N}\!\big(q_p;\,q_g(\rho_p;\boldsymbol{\theta}_g^{(t)}),\,\sigma_q^2\big)}
		{\sum\limits_{g'\in\mathcal{G}}\pi_{g'}^{(t)}(x_p)\,\mathcal{N}\!\big(q_p;\,q_{g'}(\rho_p;\boldsymbol{\theta}_{g'}^{(t)}),\,\sigma_q^2\big)}.
	\end{aligned}
	\label{eq:responsibility}
\end{equation}
Gaussian kernel aggregation then converts these point-wise responsibilities into a continuous spatial target mixture,
\begin{equation}
	\begin{aligned}
		\pi_{g,\mathrm{tar}}^{(t)}(x)
		&=\frac{\sum\limits_{p=1}^N w_p\,K(x,x_p;\sigma)\,\gamma_{p,g}^{(t)}}
		{\sum\limits_{p=1}^N w_p\,K(x,x_p;\sigma)},\\
		K(x,x_p;\sigma)
		&=\exp\!\left(-\frac{(x-x_p)^2}{2\sigma^2}\right),
	\end{aligned}
	\label{eq:kernel_aggregation}
\end{equation}
weighted by the quasi-stationary score $w_p$.

\subsection{Spatial Aggregation and Regularized Spin Update}
The M-step updates the prototype parameters via weighted least squares,
\begin{equation}
	\boldsymbol{\theta}_g^{(t+1)}=\arg\min_{\boldsymbol{\theta}_g}\sum_{p=1}^N \gamma_{p,g}^{(t)}\,w_p\Big(q_g(\rho_p;\boldsymbol{\theta}_g)-q_p\Big)^2,
	\label{eq:mstep_theta}
\end{equation}
subject to feasibility constraints in Sec.~II.
Simultaneously, the spin field update aligns the induced mixture $\boldsymbol{\pi}(x)=\Pi(\mathbf{s}(x))$ with the aggregated target $\boldsymbol{\pi}_{\mathrm{tar}}^{(t)}(x)$. 
The spin field $\mathbf{s}(x)$ evolves to minimize a regularized free energy functional that balances data consistency against physical conservation laws:
\begin{equation}
	\begin{aligned}
		\mathbf{s}^{(t+1)}&=\arg\min_{\mathbf{s}} \int_0^L \Big[ -\sum_{g\in\mathcal{G}} \pi_{g,\mathrm{tar}}^{(t)}(x)\log \pi_g(x)\\
		&\quad +\lambda_{\mathrm{phys}}(\partial_t\rho+\partial_x q)^2 +\lambda_{\mathrm{smooth}}\|\partial_x\mathbf{s}\|_2^2 \Big]dx.
	\end{aligned}
	\label{eq:spin_objective}
\end{equation}
where $q(\rho;x)$ denotes the mixture FD. 
Phase Unfolding lets $|\mathbf{s}(x)|$ grow where phase evidence is sharp and stay small in mixed regimes (no fixed radius on $\|\mathbf{s}\|$). The complete procedure is summarized in Algorithm~\ref{alg:spinflow_em}.

\begin{algorithm}[t]
	\caption{EM-Based Joint Spin--FD Inference}
	\label{alg:spinflow_em}
	{\small
	\begin{algorithmic}[1]
		\Require Observations $\{(\rho_p,q_p,x_p,w_p)\}_{p=1}^{N}$; prototype FDs $\{q_g(\cdot;\boldsymbol{\theta}_g)\}_{g\in\mathcal{G}}$; grid $\{x_i\}_{i=1}^{M}$
		\Require Hyperparameters $\lambda_{\mathrm{phys}},\lambda_{\mathrm{smooth}},\sigma,\beta$
		\Ensure Spin field $\mathbf{s}(x_i)$ and mixture weights $\boldsymbol{\pi}(x_i)$
		\State Initialize $\boldsymbol{\theta}_g^{(0)}$ and $\mathbf{s}^{(0)}(x_i)$
		\For{$t=0,1,\ldots$ until convergence}
			\State $\boldsymbol{\pi}^{(t)}(x_i)\leftarrow\Pi\!\bigl(\mathbf{s}^{(t)}(x_i)\bigr)$ via \eqref{eq:softmax_mapping}
			\State \textbf{E-step:} compute $\gamma_{p,g}^{(t)}$ via \eqref{eq:responsibility}
			\State aggregate $\pi_{g,\mathrm{tar}}^{(t)}(x_i)$ via \eqref{eq:kernel_aggregation}
			\State \textbf{M-step (FD):} update $\boldsymbol{\theta}_g^{(t+1)}$ via \eqref{eq:mstep_theta}
			\State \textbf{M-step (spin):} update $\mathbf{s}^{(t+1)}$ by minimizing \eqref{eq:spin_objective}
		\EndFor
		\State \Return $\mathbf{s}(x_i)$ and $\boldsymbol{\pi}(x_i)$
	\end{algorithmic}
	}
\end{algorithm}

\subsection{Phase Equilibrium Degree and Transition Detection}
We quantify local mixture equilibrium by the mismatch between the EM-aggregated target $\boldsymbol{\pi}_{\mathrm{tar}}(x)$ and the inferred mixture $\boldsymbol{\pi}(x)$. 
Defining the free-energy gap as
\begin{equation}
	\begin{aligned}
		\Delta\mathcal{F}(x)
		&=D_{\mathrm{KL}}\!\big(\boldsymbol{\pi}_{\mathrm{tar}}(x)\,\|\,\boldsymbol{\pi}(x)\big)\\
		&=\sum_{g\in\mathcal{G}} \pi_{g,\mathrm{tar}}(x)\log\frac{\pi_{g,\mathrm{tar}}(x)}{\pi_g(x)},
	\end{aligned}
	\label{eq:free_energy_gap}
\end{equation}
we introduce the Phase Equilibrium Degree (PED):
\begin{equation}
	\mathrm{PED}(x)=\exp\!\big(-\Delta\mathcal{F}(x)\big)\in(0,1].
	\label{eq:ped}
\end{equation}
The phase-coexistence window $\Omega_Q$ is formed by thresholding the Shannon entropy $H(x)$ of $\boldsymbol{\pi}(x)$:
\begin{equation}
	\Omega_Q=\{x \mid H(x)\ge\tau_H=\ln 2\},
	\label{eq:omega_q}
\end{equation}
where $\tau_H=\ln 2$ is the two-phase equal-weight boundary.
The \emph{primary bottleneck site} is then defined as
\begin{equation}
	x^{\ast}=\arg\min_{x\in\Omega_Q}\,\mathrm{PED}(x),
	\label{eq:xstar}
\end{equation}
the cell at which the model–data equilibrium is most severely broken. 
$x^\ast$ identifies the nucleation center of the transition and represents the highest-priority location for ATM intervention. The full detection procedure is given in Algorithm~\ref{alg:ped_detection}.

\begin{algorithm}[t]
	\caption{PED-Based Phase-Transition Detection}
	\label{alg:ped_detection}
	{\small
	\begin{algorithmic}[1]
		\Require $\boldsymbol{\pi}(x_i)$, observed $q_\mathrm{obs}$, $\rho_\mathrm{obs}$ from Alg.~\ref{alg:spinflow_em}; threshold $\tau_H{=}\ln 2$; margin $\delta$
		\Ensure Primary site $x^\ast$; profile $\{\mathrm{PED}(x_i)\}$; metric $\Delta\mathrm{PED}$
		\State $\forall\,x_i$: $H(x_i)\leftarrow -\sum_{g\in\mathcal{G}}\pi_g\log\pi_g$
		\Statex \phantom{$\forall\,x_i$:}\,and $\mathrm{PED}(x_i)\leftarrow\exp\!\bigl(-\Delta\mathcal{F}(x_i)\bigr)$ via \eqref{eq:free_energy_gap}--\eqref{eq:ped}
		\State $\Omega_Q\leftarrow\{x_i\in[\delta,L{-}\delta]\mid H(x_i)\ge\tau_H\}$ via \eqref{eq:omega_q}
		\If{$\Omega_Q\ne\emptyset$}
			\State $x^\ast\leftarrow\arg\min_{x\in\Omega_Q}\mathrm{PED}(x)$ via \eqref{eq:xstar}
		\Else
			\State $x^\ast\leftarrow\arg\max_{x\in[\delta,\,L-\delta]}|\nabla_x\rho_\mathrm{obs}|$ \Comment{density-gradient fallback}
		\EndIf
		\State $\Delta\mathrm{PED}\leftarrow 1-\mathrm{PED}(x^\ast)\big/\mathrm{mean}\bigl\{\mathrm{PED}(x_i)\mid x_i\notin\Omega_Q\bigr\}$
		\State \Return $x^\ast$, $\{\mathrm{PED}(x_i)\}$, $\Delta\mathrm{PED}$
	\end{algorithmic}
	}
\end{algorithm}

\section{Experimental Validation}
\label{sec:experiments}

\subsection{Datasets and Setup}
\label{subsec:setup}

\begin{figure}[t]
\centering
\includegraphics[width=\columnwidth]{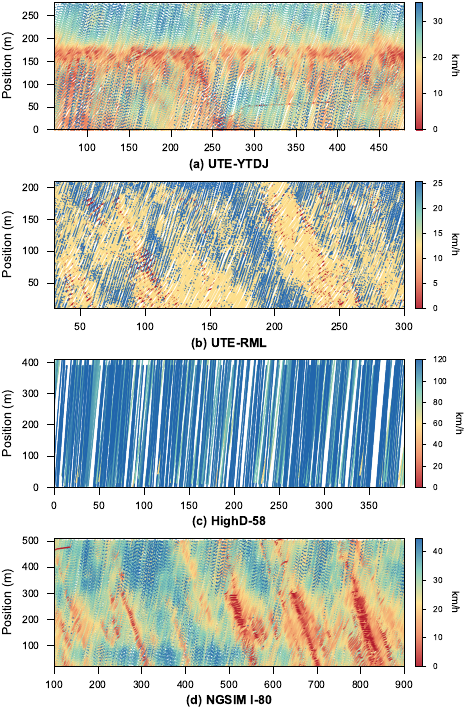}
\caption{Observed spacetime speed diagrams synthesized by projecting all analyzed-lane vehicle trajectories onto the time--position plane.}
\label{fig:spacetime_observed}
\end{figure}

Validation uses four real-world trajectory datasets: YTDJ and RML from the Ubiquitous Traffic Eyes (UTE, China) \cite{feng2025ubiquitous}, HighD-58 (Germany) \cite{krajewski2018highd}, and NGSIM I-80 (USA) \cite{coifman2017ngsim}, covering expressway bottlenecks, on-ramps, and a highway section. All datasets provide per-vehicle position and speed at dataset-specific sampling rates; Fig.~\ref{fig:spacetime_observed} shows their observed spacetime speed patterns. Table~\ref{tab:datasets} gives key statistics.

\begin{table}[h]
\caption{Dataset statistics. FPS: sampling rate; $N_\mathrm{veh}$: vehicles in the analysis window; $\bar{v}$: mean speed; FD pts: sampled observations.}
\label{tab:datasets}
\centering\renewcommand{\arraystretch}{1.15}\footnotesize
\resizebox{\columnwidth}{!}{%
\begin{tabular}{l c c c c c c c}
\toprule
Scenario & Length (m) & Duration (s) & FPS & $N_\mathrm{veh}$ & Lanes & $\bar{v}$ (km/h) & FD pts \\
\midrule
UTE-YTDJ   & 362 & 545 & 24 &  1072 & 5 & 17.2 & 512 \\
UTE-RML    & 220 & 300 & 30 &  555 & 4 & 15.2 & 564 \\
HighD-58     & 425 & 389 & 25 &  373 & 4 & 31.0 & 161 \\
NGSIM I-80   & 540 & 800 & 10 & 1634 & 5 & 25.0 & 232 \\
\bottomrule
\end{tabular}%
}
\end{table}

For each scenario, FD points are sampled following Sec.~II. The grid uses $\Delta x=2$\,m. All models share the same FD points, spatial grid, and seeds. Experiments are repeated over five seeds (42, 137, 256, 314, 628). Statistical significance is assessed by paired $t$-tests on RMSE$_q$ (SpinFlow vs.\ alternatives, two-sided). Key hyperparameters: $\lambda_{\mathrm{phys}}=0.1$, $\lambda_{\mathrm{smooth}}=0.02$, $\lambda_{\mathrm{fd}}=1.0$, learning rate~$=0.05$, EM iterations~$=80$, inner steps~$=20$, tolerance $5\times10^{-4}$. Sensitivity analysis is provided in the companion repository.

\subsection{Evaluation Metrics and Protocol}
\label{subsec:metrics}

\textbf{Forward consistency} evaluates how the inferred mixture reproduces macroscopic observations. Mixture-implied flow and speed for observation $p$ follow the mixture $q(\rho_p;x_p)$ and $v(\rho_p;x_p)$ (Sec.~III; $v$ at $\rho_p=0$ is the weighted free-flow speed, see~\eqref{eq:mixture_speed}). We report RMSE$_q$, $R_q^2$, RMSE$_v$, and $R_v^2$, where lower RMSE and higher $R^2$ indicate better fit.

\textbf{Transition detection} uses the PED-based bottleneck localization from Sec.~IV-C. We report the primary bottleneck site $x^\ast$, the $\mathrm{argmin}$ of $\mathrm{PED}$ over $\Omega_Q$, and its transition mean absolute error, T.MAE, against a reference location, together with $\mathrm{PED}_\mathrm{min}$, $\Delta\mathrm{PED}$ as the relative drop from the stable-zone baseline, and $H(x^\ast)$, the Shannon entropy at $x^\ast$.

\textbf{Interpretability and physics consistency.} Phase entropy standard deviation $\sigma_H$ measures how phase-mixing entropy $H(x)$ varies along the segment coordinate $x$; it is not monotonic as a figure of merit. Physical residual (Phys.\,Res.) is the mean per cell of $(\partial_t\rho + \partial_x q)^2$ on the inferred mixture; lower values indicate better physics consistency.

\subsection{Main Results and Interpretation}
\label{subsec:main_results}

\begin{figure*}[h]
\centering
\includegraphics[width=1\textwidth]{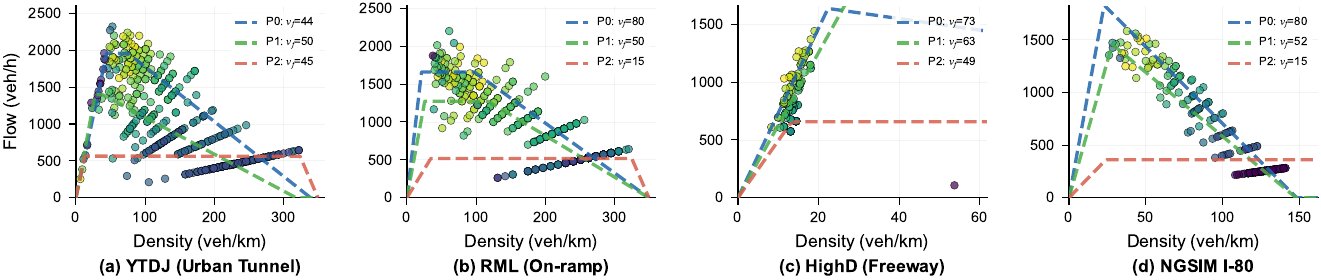}
\caption{Fitted three-phase prototype FDs for all four scenarios. Scatter points are parallelogram-sampled observations colored by quality score; dashed curves show the calibrated triangular FD prototypes (F, S, J).}
\label{fig:fd_prototypes}
\end{figure*}

Fig.~\ref{fig:fd_prototypes} presents the three EM-calibrated prototype FDs. Quality-weighted observations cluster tightly around each triangular prototype across all scenarios, validating the parallelogram sampler's efficacy in isolating quasi-stationary states.

\begin{figure*}[h]
\centering
\includegraphics[width=1\textwidth]{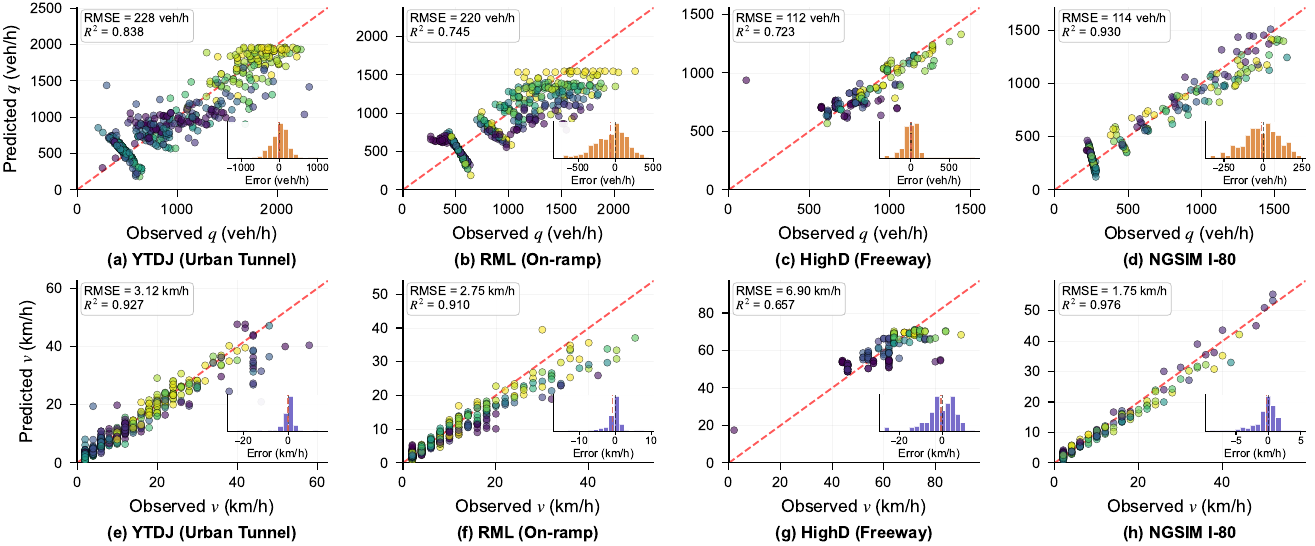}
\caption{Forward consistency. Top row~(a--d): observed vs.\ predicted flow~$q$ for all four scenarios. Bottom row~(e--h): observed vs.\ predicted speed~$v$. Points are colored by spatial position~$x$; insets show residual histograms.}
\label{fig:forward_consistency}
\end{figure*}

Fig.~\ref{fig:forward_consistency} confirms a tight 1:1 clustering of reconstructed flow and speed with near-zero residuals. Overall, SpinFlow achieves $R_q^2 \geq 0.72$ and $R_v^2 \geq 0.65$ universally (Table~\ref{tab:all_perf}). Performance peaks on the balanced NGSIM I-80 dataset ($R_q^2 = 0.940, R_v^2 = 0.981$), remains robust under heavy urban congestion for YTDJ and RML ($R_q^2 = 0.837$ and $0.744$), and holds steady on sparse HighD data ($R_q^2 = 0.724$). Computationally, the EM algorithm converges within 23--79 iterations, successfully reducing total loss by 96\%, even when reaching the 80-iter ceiling on NGSIM.
EM convergence curves are also provided in the companion repository.

\begin{figure*}[h]
\centering
\includegraphics[width=1\textwidth]{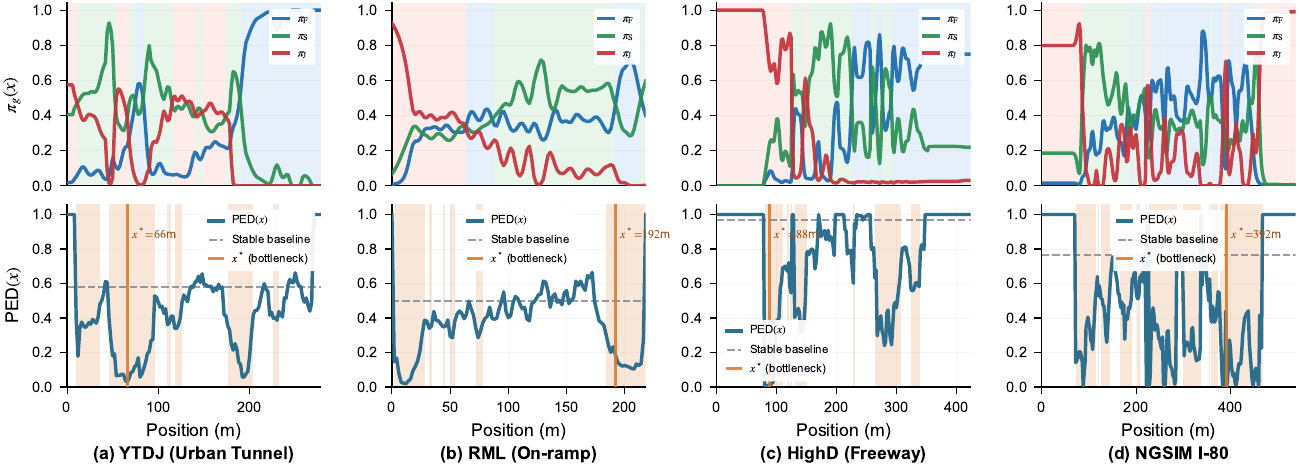}
\caption{Phase inference and bottleneck localization for all four scenarios. Top: phase weights $\boldsymbol{\pi}_g(x)$ (blue\,=\,F, green\,=\,S, pink\,=\,J). Bottom: PED$(x)$ with non-equilibrium zones (red fill), stable baseline (dashed), and $x^\ast$ (red line).}
\label{fig:interpretability}
\end{figure*}

Fig.~\ref{fig:interpretability} and Table~\ref{tab:transition} detail inferred phase profiles and PED-based detections. Severe PED drops (94.9--100\%) together with $H(x^\ast) > \ln 2$ are consistent with mixed-phase coexistence at $x^\ast$. Reported $x^\ast$ locations agree with documented bottleneck features on YTDJ ($x^\ast = 66$\,m, upstream lane-drop), RML ($x^\ast = 192$\,m, on-ramp mixing edge), and NGSIM I-80 ($x^\ast = 392$\,m, downstream on-ramp boundary). HighD gives the smallest $\mathrm{PED}_\mathrm{min}=0.001$ at $x^\ast = 88$\,m.

\begin{table}[h]
\caption{Bottleneck detection. $x^\ast$: primary site; $\Delta\mathrm{PED}$: drop vs.\ stable baseline; $H(x^\ast)$: entropy at $x^\ast$.}
\label{tab:transition}
\centering\renewcommand{\arraystretch}{1.15}\footnotesize
\resizebox{\columnwidth}{!}{%
\begin{tabular}{l c c c c c}
\toprule
Scenario & $x^\ast$ (m) & PED$_\mathrm{min}$ & PED$_\mathrm{st}$ & $\Delta\mathrm{PED}$ (\%) & $H(x^\ast)$ \\
\midrule
YTDJ (UTE)  &  66 & 0.030 & 0.577 & 94.9 & 1.071 \\
RML (UTE)   & 192 & 0.019 & 0.499 & 96.3 & 0.912 \\
HighD-58    &  88 & 0.001 & 0.971 & 99.9 & 0.809 \\
NGSIM I-80  & 392 & 0.000 & 0.763 & 100.0 & 0.943 \\
\bottomrule
\end{tabular}%
}
\end{table}

\subsection{Baseline and Ablation Analysis}
\label{subsec:comparison}

We compare SpinFlow against three heterogeneous baselines representing different modeling paradigms. Table~\ref{tab:all_perf} details their performance and runtime over 5 seeds.\\
\textbf{PWA-CTM (White-box):} Inspired by piecewise-affine approximations of the CTM \cite{alimardani2022pwa}, this original baseline blends three triangular FD zones via an adaptive sigmoid function to explicitly simulate gradual phase coexistence.\\
\textbf{VBGMM+KDE (Grey-box):} Building on GMM-based traffic classification \cite{liu2016real}, this applies a variational Bayesian Gaussian mixture in $(\rho,q,x)$ space. It incorporates Gaussian KDE for spatial smoothing, enabling continuous phase weight fitting.\\
\textbf{PI-DeepONet (Black-box):} Based on physics-informed operator learning for traffic \cite{li2025physics}, this Branch--Trunk network is trained with MSE and an LWR continuity loss. It predicts pointwise flow but offers no phase weights or PED signals.

\textbf{Baseline results.}
SpinFlow achieves the lowest RMSE$_q$ on YTDJ, HighD, and NGSIM ($p{<}0.05$ vs.\ PWA-CTM and VBGMM+KDE; by the same paired RMSE$_q$ test, PI-DeepONet lags SpinFlow on YTDJ and NGSIM). On near-homogeneous RML, PI-DeepONet attains lower RMSE$_q$ (201.8 vs.\ 219.8\,veh/h) but outputs neither calibrated $\boldsymbol{\pi}(x)$ nor PED. 
On YTDJ, PWA-CTM is constrained when coexistence is gradual (327.6\,veh/h), whereas PI-DeepONet degrades strongly ($R_v^2{=}{-}0.42$) under strong LWR continuity regularization. On HighD, SpinFlow attains T.MAE${=}0.0$\,m with RMSE$_q{=}111.9$\,veh/h; PI-DeepONet yields comparable flow RMSE on average ($p{=}0.055$) but larger variance across seeds. On NGSIM, SpinFlow reaches $R_q^2{=}0.940$.

\begin{table*}[h]
\caption{Accuracy and efficiency across four scenarios. Best results in bold; $^\ast$: $p{<}0.05$ vs.\ SpinFlow (paired $t$-test on RMSE$_q$).}
\label{tab:all_perf}
\centering\renewcommand{\arraystretch}{1.08}\scriptsize
\resizebox{\textwidth}{!}{%
\begin{tabular}{l l c c c c c c c c r c}
\toprule
Scenario & Model & RMSE$_q$\,$\downarrow$ & $R_q^2$\,$\uparrow$ & RMSE$_v$\,$\downarrow$ & $R_v^2$\,$\uparrow$ & T.MAE\,(m)\,$\downarrow$ & Phys.\,Res.\,$\downarrow$ & $\sigma_H$\,$\leftrightarrow$ & RT\,(s)\,$\downarrow$ & Iter\,$\downarrow$ & $\pi(x)$ \\
\midrule
\multirow{8}{*}{\makecell[l]{UTE-YTDJ\\(Bottleneck)}}
 & \cellO{SpinFlow}            & \cellO{\textbf{228.1}\std{0.0}} & \cellO{\textbf{0.837}} & \cellO{\textbf{3.14}\std{0.00}} & \cellO{\textbf{0.926}} & \cellO{34.0\std{0.0}}   & \cellO{5.29e{-4}}   & \cellO{0.355} & \cellO{3.61\std{0.66}} & \cellO{59}   & \cellO{Yes}     \\
 & PWA-CTM$^\ast$      & 327.6\std{0.3}         & 0.664          & 3.42\std{0.00}         & 0.913 & 28.0\std{0.0}   & \textbf{2.84e{-4}} & 0.238 & 0.07\std{0.00} & \textbf{1}    & Partial \\
 & VBGMM+KDE$^\ast$    & 277.8\std{0.0}         & 0.759          & 3.83\std{0.00}         & 0.890 & 114.0\std{0.0}  & 4.63e{-4}   & 0.282 & 0.67\std{0.78} & 22   & Partial \\
 & PI-DeepONet$^\ast$  & 361.9\std{29.9}        & 0.588          & 13.69\std{1.80}        & $-$0.418 & \textbf{0.0}\std{0.0} & 3.61e{-4}   & 0.000 & 5.25\std{2.38} & 1420 & No      \\
\cmidrule{2-12}
 & \cellAb{100}{No Comp.$^\ast$}  & \cellAb{100}{241.0\std{0.1}} & \cellAb{100}{0.818} & \cellAb{100}{3.21\std{0.00}} & \cellAb{100}{0.917} & \cellAb{100}{34.0\std{0.0}}   & \cellAb{100}{4.93e{-4}} & \cellAb{100}{0.234} & \cellAb{100}{1.97\std{0.09}} & \cellAb{100}{36} & \cellAb{100}{Yes} \\
 & \cellAb{72}{Unit Norm$^\ast$} & \cellAb{72}{239.8\std{1.3}} & \cellAb{72}{0.820} & \cellAb{72}{3.30\std{0.03}} & \cellAb{72}{0.919} & \cellAb{72}{34.0\std{0.0}}   & \cellAb{72}{4.53e{-4}} & \cellAb{72}{0.273} & \cellAb{72}{1.60\std{0.32}} & \cellAb{72}{25} & \cellAb{72}{Yes} \\
 & \cellAb{48}{No Phys.$^\ast$}  & \cellAb{48}{230.6\std{0.0}} & \cellAb{48}{0.834} & \cellAb{48}{3.26\std{0.00}} & \cellAb{48}{0.921} & \cellAb{48}{34.0\std{0.0}}   & \cellAb{48}{5.12e{-4}} & \cellAb{48}{0.350} & \cellAb{48}{2.25\std{0.18}} & \cellAb{48}{42} & \cellAb{48}{Yes} \\
 & \cellAb{26}{Single FD$^\ast$} & \cellAb{26}{372.8\std{0.0}} & \cellAb{26}{0.565} & \cellAb{26}{3.90\std{0.00}} & \cellAb{26}{0.886} & \cellAb{26}{\textbf{0.0}\std{0.0}} & \cellAb{26}{3.28e{-4}} & \cellAb{26}{0.000} & \cellAb{26}{\textbf{0.01}\std{0.00}} & \cellAb{26}{\textbf{1}}  & \cellAb{26}{Yes} \\
\midrule
\multirow{8}{*}{\makecell[l]{UTE-RML\\(On-ramp)}}
 & \cellO{SpinFlow}            & \cellO{219.8\std{0.0}}         & \cellO{0.744}          & \cellO{2.76\std{0.00}} & \cellO{0.909} & \cellO{166.0\std{0.0}}  & \cellO{2.38e{-4}}   & \cellO{0.184} & \cellO{2.75\std{0.06}} & \cellO{75}   & \cellO{Yes}     \\
 & PWA-CTM$^\ast$      & 224.5\std{0.1}         & 0.733          & 1.99\std{0.01} & 0.953 & 122.0\std{0.0}  & 3.63e{-4}   & 0.245 & \textbf{0.05}\std{0.00} & \textbf{1}    & Partial \\
 & VBGMM+KDE$^\ast$    & 381.7\std{0.1}         & 0.228          & 4.82\std{0.01} & 0.723 & 166.0\std{0.0}  & \textbf{2.09e{-4}} & 0.223 & 0.45\std{0.06} & 46   & Partial \\
 & PI-DeepONet$^\ast$  & \textbf{201.8}\std{1.2} & \textbf{0.784} & \textbf{1.89}\std{0.01} & \textbf{0.957} & \textbf{98.0}\std{0.0} & 4.30e{-4}   & 0.271 & 4.50\std{1.03} & 1321 & No      \\
\cmidrule{2-12}
 & \cellAb{100}{No Comp.$^\ast$}  & \cellAb{100}{234.1\std{0.0}} & \cellAb{100}{0.710} & \cellAb{100}{3.03\std{0.00}} & \cellAb{100}{0.890} & \cellAb{100}{166.0\std{0.0}}  & \cellAb{100}{2.22e{-4}} & \cellAb{100}{0.110} & \cellAb{100}{0.78\std{0.01}} & \cellAb{100}{17} & \cellAb{100}{Yes} \\
 & \cellAb{72}{Unit Norm$^\ast$} & \cellAb{72}{230.9\std{0.5}} & \cellAb{72}{0.718} & \cellAb{72}{3.03\std{0.01}} & \cellAb{72}{0.890} & \cellAb{72}{166.0\std{0.0}}  & \cellAb{72}{2.16e{-4}} & \cellAb{72}{0.127} & \cellAb{72}{1.59\std{0.33}} & \cellAb{72}{40} & \cellAb{72}{Yes} \\
 & \cellAb{48}{No Phys.$^\ast$}  & \cellAb{48}{231.0\std{0.0}} & \cellAb{48}{0.718} & \cellAb{48}{3.04\std{0.00}} & \cellAb{48}{0.890} & \cellAb{48}{166.0\std{0.0}}  & \cellAb{48}{2.25e{-4}} & \cellAb{48}{0.151} & \cellAb{48}{1.51\std{0.03}} & \cellAb{48}{39} & \cellAb{48}{Yes} \\
 & \cellAb{26}{Single FD$^\ast$} & \cellAb{26}{224.5\std{0.0}} & \cellAb{26}{0.733} & \cellAb{26}{1.95\std{0.00}} & \cellAb{26}{0.955} & \cellAb{26}{\textbf{98.0}\std{0.0}} & \cellAb{26}{3.18e{-4}} & \cellAb{26}{0.000} & \cellAb{26}{\textbf{0.02}\std{0.00}} & \cellAb{26}{\textbf{1}}  & \cellAb{26}{Yes} \\
\midrule
\multirow{8}{*}{\makecell[l]{HighD-58\\(Segment)}}
 & \cellO{SpinFlow}            & \cellO{\textbf{111.9}\std{0.0}} & \cellO{\textbf{0.724}} & \cellO{\textbf{6.89}\std{0.00}} & \cellO{\textbf{0.658}} & \cellO{\textbf{0.0}\std{0.0}} & \cellO{1.25e{-5}}   & \cellO{0.279} & \cellO{0.70\std{0.02}} & \cellO{23}   & \cellO{Yes}     \\
 & PWA-CTM$^\ast$      & 133.5\std{0.7}         & 0.608          & 8.13\std{0.06}         & 0.525 & 127.8\std{58.0} & 1.33e{-5}   & 0.208 & 0.05\std{0.00} & \textbf{1}    & Partial \\
 & VBGMM+KDE$^\ast$    & 138.4\std{0.0}         & 0.578          & 8.55\std{0.00}         & 0.474 & 6.0\std{0.0}    & 1.15e{-5}   & 0.434 & 0.27\std{0.10} & 36   & Partial \\
 & PI-DeepONet         & 143.5\std{25.6}        & 0.535          & 9.90\std{1.14}         & 0.287 & 119.8\std{0.0}  & 1.40e{-5}   & 0.013 & 6.56\std{0.05} & 2000 & No      \\
\cmidrule{2-12}
 & \cellAb{100}{No Comp.$^\ast$}  & \cellAb{100}{136.0\std{0.7}}          & \cellAb{100}{0.593}          & \cellAb{100}{8.00\std{0.03}}          & \cellAb{100}{0.539} & \cellAb{100}{6.0\std{0.0}}   & \cellAb{100}{1.13e{-5}} & \cellAb{100}{0.214} & \cellAb{100}{0.68\std{0.08}} & \cellAb{100}{21} & \cellAb{100}{Yes} \\
 & \cellAb{72}{Unit Norm$^\ast$} & \cellAb{72}{118.8\std{0.1}}          & \cellAb{72}{0.689}          & \cellAb{72}{7.39\std{0.01}}          & \cellAb{72}{0.607} & \cellAb{72}{2.0\std{0.0}}   & \cellAb{72}{1.22e{-5}} & \cellAb{72}{0.218} & \cellAb{72}{0.66\std{0.19}} & \cellAb{72}{21} & \cellAb{72}{Yes} \\
 & \cellAb{48}{No Phys.$^\ast$}  & \cellAb{48}{112.7\std{0.0}} & \cellAb{48}{0.720} & \cellAb{48}{6.93\std{0.00}} & \cellAb{48}{0.655} & \cellAb{48}{2.0\std{0.0}}   & \cellAb{48}{1.25e{-5}} & \cellAb{48}{0.270} & \cellAb{48}{2.25\std{0.03}} & \cellAb{48}{80} & \cellAb{48}{Yes} \\
 & \cellAb{26}{Single FD$^\ast$} & \cellAb{26}{171.0\std{0.0}}          & \cellAb{26}{0.355}          & \cellAb{26}{10.97\std{0.00}}         & \cellAb{26}{0.134} & \cellAb{26}{119.8\std{0.0}} & \cellAb{26}{\textbf{1.09e{-5}}} & \cellAb{26}{0.000} & \cellAb{26}{\textbf{0.02}\std{0.00}} & \cellAb{26}{\textbf{1}}  & \cellAb{26}{Yes} \\
\midrule
\multirow{8}{*}{\makecell[l]{NGSIM I-80\\(Bottleneck)}}
 & \cellO{SpinFlow}            & \cellO{\textbf{105.7}\std{1.3}} & \cellO{\textbf{0.940}} & \cellO{\textbf{1.53}\std{0.05}}         & \cellO{\textbf{0.981}} & \cellO{308.0\std{0.0}}  & \cellO{7.30e{-5}}   & \cellO{0.298} & \cellO{3.94\std{0.08}} & \cellO{79}   & \cellO{Yes}     \\
 & PWA-CTM$^\ast$      & 123.3\std{2.4}         & 0.918          & 1.59\std{0.03}         & 0.980 & \textbf{201.2}\std{1.1} & 8.14e{-5}   & 0.221 & 0.06\std{0.00} & \textbf{1}    & Partial \\
 & VBGMM+KDE$^\ast$    & 270.3\std{0.1}         & 0.606          & 5.45\std{0.00}         & 0.763 & 300.0\std{0.0}  & 1.43e{-4}   & 0.299 & 0.33\std{0.07} & 31   & Partial \\
 & PI-DeepONet$^\ast$  & 118.9\std{4.7}         & 0.924          & 1.76\std{0.15}         & 0.975 & 304.0\std{0.0}  & 1.10e{-4}   & 0.170 & 5.30\std{2.13} & 1560 & No      \\
\cmidrule{2-12}
 & \cellAb{100}{No Comp.$^\ast$}  & \cellAb{100}{131.4\std{0.0}} & \cellAb{100}{0.907} & \cellAb{100}{2.08\std{0.00}} & \cellAb{100}{0.966} & \cellAb{100}{320.0\std{0.0}} & \cellAb{100}{\textbf{5.23e{-5}}} & \cellAb{100}{0.255} & \cellAb{100}{1.32\std{0.05}} & \cellAb{100}{23} & \cellAb{100}{Yes} \\
 & \cellAb{72}{Unit Norm$^\ast$} & \cellAb{72}{130.9\std{1.3}} & \cellAb{72}{0.908} & \cellAb{72}{2.11\std{0.03}} & \cellAb{72}{0.964} & \cellAb{72}{308.0\std{0.0}} & \cellAb{72}{5.65e{-5}}  & \cellAb{72}{0.229} & \cellAb{72}{2.20\std{1.66}} & \cellAb{72}{41} & \cellAb{72}{Yes} \\
 & \cellAb{48}{No Phys.$^\ast$}  & \cellAb{48}{130.3\std{0.2}} & \cellAb{48}{0.908} & \cellAb{48}{2.07\std{0.01}} & \cellAb{48}{0.966} & \cellAb{48}{308.0\std{0.0}} & \cellAb{48}{5.47e{-5}}  & \cellAb{48}{0.320} & \cellAb{48}{1.05\std{0.21}} & \cellAb{48}{16} & \cellAb{48}{Yes} \\
 & \cellAb{26}{Single FD$^\ast$} & \cellAb{26}{121.9\std{0.0}} & \cellAb{26}{0.920} & \cellAb{26}{1.55\std{0.00}} & \cellAb{26}{0.978} & \cellAb{26}{304.0\std{0.0}} & \cellAb{26}{7.85e{-5}}  & \cellAb{26}{0.000} & \cellAb{26}{\textbf{0.02}\std{0.00}} & \cellAb{26}{\textbf{1}}  & \cellAb{26}{Yes} \\
\midrule
\multicolumn{12}{@{}l@{}}{\footnotesize Mean$\pm$std over 5 seeds. Units: RMSE$_q$\,(veh/h), RMSE$_v$\,(km/h), Phys.\,Res.\,$(\mathrm{veh/(m\cdot s)})^2$. $\pi(x)$: Yes\,=\,calibrated, Partial\,=\,proxy, No\,=\,N/A.} \\
\bottomrule
\end{tabular}%
}
\end{table*}

\textbf{Ablation results.}
Four ablations remove one SpinFlow mechanism at a time: \textbf{No Comp.}\ disables the competition penalty ($h_S{=}s_x$); \textbf{Unit Norm}\ projects $\mathbf{s}$ onto the unit sphere, eliminating Phase Unfolding; \textbf{No Phys.}\ sets $\lambda_{\mathrm{phys}}{=}0$; \textbf{Single FD}\ reduces the model to standard single-prototype LWR.

Removing the competition mapping raises RMSE$_q$ by 5.7--24.3\,\% across all scenarios ($p{<}0.05$). On YTDJ, $\sigma_H$ decreases from 0.355 to 0.234, consistent with weaker along-$x$ variability in mixing entropy when the three-phase competition pathway is disabled; $\sigma_H$ is reported descriptively and is not used as a scalar objective. Synchronized weight $\pi_S$ is suppressed, yielding an effective free--jam partition.
Constraining the spin to unit norm increases RMSE$_q$ by 5--24\,\%, with the largest gap on NGSIM under extreme densities, indicating that Phase Unfolding helps localize sharp phase dominance.
Omitting the physics prior increases RMSE$_q$ most on NGSIM (23.2\,\%) and only modestly on YTDJ ($+$1.1\,\%); on sparse HighD the effect is negligible as LWR gradients shrink at convergence.
Single FD sharply raises RMSE$_q$ (63.4\,\% on YTDJ, 52.8\,\% on HighD) and ties transitions to a density-gradient heuristic, yielding scattered T.MAE (0, 98, 120, and 304\,m) when gradient peaks misalign with the reference bottleneck. PED sidesteps that coupling---notably T.MAE${=}0.0$\,m on HighD---without hand-tuned thresholds.

In summary, SpinFlow strikes a practical balance among accuracy, interpretability, and computational efficiency. It achieves strong in-sample flow fit ($R_q^2{\geq}0.72$), yields structured phase maps ($\sigma_H{>}0$), and supports PED-based bottleneck localization. Per scenario on a single CPU, training converges within 0.7--3.9\,s and is up to $9\times$ faster than PI-DeepONet (4.5--6.6\,s); conversely, analytic calibrations run in under $<0.1$\,s but do not perform explicit phase inference.

	\section{Conclusion}
	\label{sec:conclusion}

SpinFlow translates Kerner's three-phase theory into a computable framework, partly addressing the two critical gaps identified in Sec.~I. First, bridging microscopic mechanisms with computable states, the competitive-equilibrium mapping lets synchronized flow emerge without pre-specified static boundaries, while physics-regularized EM and Phase Unfolding yield interpretable phase maps. Second, replacing rigid empirical thresholds, PED quantifies model--data structural alignment and localizes coexistence in a topology-free manner. Across four real-world datasets, SpinFlow achieves $R_q^2$ up to 0.940 and PED drops of 94.9--100\,\%; relative to heterogeneous baselines, it delivers calibrated $\boldsymbol{\pi}(x)$ with favorable RMSE$_q$ except against PI-DeepONet on RML. SpinFlow thereby offers an ATM-oriented precursor signal grounded in phase structure rather than fixed thresholds alone. This motivates viewing traffic dynamics as spatial phase competition with emergent signatures, complementary to hydrodynamic models.
	
Limitations remain. First, gains over a single-phase FD shrink to ${\sim}2\,\%$ on near-homogeneous scenarios (e.g.\ RML), suggesting mild redundancy when multi-phase structure is weak. Second, quasi-stationary parallelogram sampling can yield spurious EM artifacts under rapidly evolving disturbances. Third, validation is confined to 1-D ordered segments: (i)~lane-changing and lateral competition are not explicitly represented; (ii)~network-scale merges and diverges, with topology-aware generalization, remain open; (iii)~online deployment calls for latency-robust incremental updates and closed-loop stability guarantees. Future work will target online incremental EM for edge-side real-time spin-field updates, multi-lane spin fields for cross-lane phase competition, graph-aware kernels with inter-segment coupling, and closed-loop integration with Vehicle-to-Everything (V2X) and Connected and Automated Vehicle (CAV) systems in which PED can trigger variable speed limits or ramp metering---advancing phase inference toward preventive ATM.

	\bibliographystyle{IEEEtran}
	\bibliography{root}

\appendices

\section{Notation Table}
\label{app:symbols}

\begin{table}[h]
\caption{Core notation used in the main text and code.}
\label{tab:symbols}
\centering\footnotesize
\renewcommand{\arraystretch}{1.12}
\resizebox{\columnwidth}{!}{%
\begin{tabular}{l l c l}
\toprule
Symbol & Meaning (unit) & Eq. & Code variable \\
\midrule
$\rho(x,t)$            & density (veh/m)                                        & \eqref{eq:edie_rqv}             & \texttt{rho\_obs} \\
$q(x,t)$               & flow (veh/s)                                           & \eqref{eq:edie_rqv}             & \texttt{q\_pred\_cells} \\
$v(x,t)$               & space-mean speed (m/s)                                 & \eqref{eq:edie_rqv}             & \texttt{v\_obs} \\
$w_p$                  & quasi-stationary weight $\in[0,1]$                     & \eqref{eq:soft_weight}          & \texttt{quality\_weights} \\
$\beta_w$              & sampler logistic sharpness                             & \eqref{eq:soft_weight}          & \texttt{beta\_w} \\
$\boldsymbol{\theta}_g$ & prototype $(v_f,w,\rho_{\mathrm{jam}},Q)$              & \eqref{eq:triangular_fd}        & \texttt{prototypes[g]} \\
$\boldsymbol{\pi}(x)$   & phase weights, $\sum_g\pi_g=1$                         & \eqref{eq:mixture_fd}           & \texttt{pi} \\
$\mathbf{s}(x)$         & latent spin vector in $\mathbb{R}^3$                   & \eqref{eq:softmax_mapping}      & \texttt{sx,sy,sz} \\
$h_g(\mathbf{s})$       & preference score for phase $g$                         & \eqref{eq:preference_scores}    & \texttt{grad\_logits} \\
$\beta$                 & Boltzmann inverse temperature                          & \eqref{eq:softmax_mapping}      & \texttt{fixed beta = 1} \\
$b(x)$                  & free--jam bias $s_z-s_y$                               & \eqref{eq:bias_strength}        & \texttt{sz - sy} \\
$\chi(x)$               & competition strength $|b|$                             & \eqref{eq:bias_strength}        & \texttt{np.abs(sz - sy)} \\
$\gamma_{p,g}$          & posterior responsibility                               & \eqref{eq:responsibility}       & \texttt{responsibilities} \\
$K(x,x_p;\sigma)$       & spatial Gaussian aggregation kernel                    & \eqref{eq:kernel_aggregation}   & \texttt{spatial\_w} \\
$\sigma$                & kernel bandwidth ($=1.5\Delta x$)                      & \eqref{eq:kernel_aggregation}   & \texttt{sigma\_spatial} \\
$\sigma_q$              & flow noise std (veh/s, data-driven)                    & \eqref{eq:flow_noise}           & \texttt{sigma\_q\_adaptive} \\
$\lambda_{\mathrm{phys}}$  & mass-conservation prior weight                      & \eqref{eq:spin_objective}       & \texttt{lam\_phys} \\
$\lambda_{\mathrm{smooth}}$ & Heisenberg-exchange prior weight                    & \eqref{eq:spin_objective}       & \texttt{lam\_smo} \\
$\lambda_{\mathrm{fd}}$ & data-fit weight                                        & \eqref{eq:mstep_theta}          & \texttt{lam\_fd\_q} \\
$\Delta\mathcal{F}(x)$  & Kullback-Leibler (KL) gap, $\pi_{\mathrm{tar}}$ vs.\ $\pi$               & \eqref{eq:free_energy_gap}     & \texttt{delta\_F} \\
$\mathrm{PED}(x)$       & Phase Equilibrium Degree $\in(0,1]$                    & \eqref{eq:ped}                 & \texttt{PED} \\
$H(x)$                  & Shannon entropy of $\boldsymbol{\pi}(x)$               & \eqref{eq:omega_q}             & \texttt{H} \\
$\tau_H$                & coexistence threshold $\ln 2$                          & \eqref{eq:omega_q}             & \texttt{tau\_H} \\
$x^{\ast}$              & primary bottleneck site (m)                            & \eqref{eq:xstar}               & \texttt{x\_star} \\
\bottomrule
\end{tabular}%
}
\end{table}

\section{Variational Free-Energy Derivation of the Physics-Regularized EM}
\label{app:em_derivation}

This appendix expands the compact statements~\eqref{eq:responsibility}--\eqref{eq:spin_objective} into a complete derivation under the variational free-energy framework of Neal and Hinton~\cite{neal1998em}, exposing the precise role of each regularizer in the M-step.

\subsection{Complete-data joint likelihood}
For each FD point $\mathbf{y}_p=(\rho_p,q_p,v_p,x_p)$ with latent phase $z_p\in\mathcal{G}$, the prototype set $\boldsymbol{\theta}=\{\boldsymbol{\theta}_g\}_{g\in\mathcal{G}}$, and the spin field $\mathbf{s}(\cdot)$, the complete-data likelihood factorizes as
\begin{equation}
\begin{aligned}
&p(q_p,z_p\mid \rho_p,x_p;\boldsymbol{\theta},\mathbf{s})\\
&\quad=\pi_{z_p}\!\bigl(x_p;\mathbf{s}\bigr)\,\mathcal{N}\!\Bigl(q_p;\,q_{z_p}\!\bigl(\rho_p;\boldsymbol{\theta}_{z_p}\bigr),\sigma_q^2\Bigr),
\end{aligned}
\label{eq:app_joint}
\end{equation}
where $\pi_g(x;\mathbf{s})$ is the softmax mapping~\eqref{eq:softmax_mapping}. The flow noise variance $\sigma_q^2$ is estimated once from prototype residuals.

\subsection{Variational Free Energy}
Introduce a variational distribution $Q_p(z)$ on the latent phase of point $p$ and define the free energy~\cite{neal1998em}
\begin{equation}
\mathcal{F}(Q,\boldsymbol{\theta},\mathbf{s})=\sum_{p=1}^{N}\!\mathbb{E}_{Q_p}\!\bigl[-\log p(q_p,z_p\mid\cdot)\bigr]-\sum_{p=1}^N\!H\!\bigl[Q_p\bigr].
\label{eq:app_F}
\end{equation}
A direct identity gives the decomposition
\begin{equation}
\mathcal{F}=-\!\sum_p\!\log p(q_p\!\mid\!\rho_p,x_p;\boldsymbol{\theta},\mathbf{s})+\!\sum_p\!D_{\mathrm{KL}}\!\bigl(Q_p\,\|\,p(z_p\!\mid\!q_p,\cdot)\bigr),
\label{eq:app_F_decomp}
\end{equation}
which exposes EM as coordinate descent on $\mathcal{F}$: the first term is the negative marginal log-likelihood; the second is a non-negative KL that the E-step zeros out.

\subsection{E-step Closed-Form Posterior}
\emph{Lemma.} For fixed $(\boldsymbol{\theta},\mathbf{s})$, the variational minimizer of~\eqref{eq:app_F} is the exact posterior
\begin{equation}
Q_p^{\star}(z\!=\!g)=p\bigl(z_p\!=\!g\,\big|\,q_p,\rho_p,x_p;\boldsymbol{\theta},\mathbf{s}\bigr)=\gamma_{p,g},
\end{equation}
which is precisely the responsibility~\eqref{eq:responsibility}. The KL term in~\eqref{eq:app_F_decomp} vanishes, so $\mathcal{F}(Q^\star,\boldsymbol{\theta},\mathbf{s})=-\sum_p\log p(q_p\!\mid\!\cdot)$.

\subsection{Augmented free energy with physical priors}
We encode physical priors on $\mathbf{s}(x)$ through log-densities
\begin{equation}
\begin{aligned}
-\log p_\mathrm{prior}(\mathbf{s})
&=\lambda_{\mathrm{phys}}\!\int_0^L\!\bigl(\partial_t\rho+\partial_x q\bigr)^2 dx\\
&\quad+\lambda_{\mathrm{smooth}}\!\int_0^L\!\|\partial_x\mathbf{s}\|_2^2\,dx+\mathrm{const.}
\end{aligned}
\label{eq:app_logprior}
\end{equation}
The first term is the LWR mass-conservation residual (Gaussian prior on PDE compliance); the second is a Heisenberg-exchange smoothness prior penalizing spatial fluctuations of the spin vector. The maximum a posteriori (MAP)-EM objective then becomes the augmented free energy
\begin{equation}
\widetilde{\mathcal{F}}\;=\;\mathcal{F}\;-\;\log p_\mathrm{prior}(\mathbf{s}),
\label{eq:app_Ftilde}
\end{equation}
which preserves the coordinate-descent structure.

\subsection{M-step on prototype parameters}
Substituting $Q^\star=\gamma$ into~\eqref{eq:app_F} and isolating $\boldsymbol{\theta}$-dependent terms yields, up to constants in $\boldsymbol{\theta}$,
\begin{equation}
\widetilde{\mathcal{F}}(\boldsymbol{\theta})=\frac{1}{2\sigma_q^2}\sum_{p,g}\gamma_{p,g}\bigl(q_g(\rho_p;\boldsymbol{\theta}_g)-q_p\bigr)^2.
\end{equation}
Phase-wise minimization recovers~\eqref{eq:mstep_theta}; the soft sampler weight $w_p$ enters multiplicatively in the implementation.

\subsection{M-step on the spin field}
After kernel aggregation~\eqref{eq:kernel_aggregation} converts pointwise responsibilities into the continuous target $\pi_{g,\mathrm{tar}}^{(t)}(x)$, the $\mathbf{s}$-dependent part of $\widetilde{\mathcal{F}}$ reads
\begin{equation}
\begin{aligned}
\widetilde{\mathcal{F}}(\mathbf{s})
&=-\!\int_0^L\!\sum_{g\in\mathcal{G}} \pi_{g,\mathrm{tar}}^{(t)}(x)\log\pi_g(x;\mathbf{s})\,dx\\
&\quad+\lambda_{\mathrm{phys}}\!\int_0^L\!(\partial_t\rho+\partial_x q)^2 dx\\
&\quad+\lambda_{\mathrm{smooth}}\!\int_0^L\!\|\partial_x\mathbf{s}\|_2^2\,dx,
\end{aligned}
\label{eq:app_spinobj}
\end{equation}

\subsection{Spin gradient via chain rule}
Define logits $\ell_g(x)=\beta\, h_g(\mathbf{s}(x))$ so that $\pi_g(x)=\mathrm{softmax}(\ell(x))_g$. The cross-entropy in~\eqref{eq:app_spinobj} yields the standard softmax-CE gradient
\begin{equation}
\frac{\partial \widetilde{\mathcal{F}}}{\partial \ell_g(x)}=\pi_g(x)-\pi_{g,\mathrm{tar}}(x).
\label{eq:app_logit_grad}
\end{equation}
Writing $\xi(x)=\mathrm{sgn}\bigl(s_z(x)-s_y(x)\bigr)$ and
$\delta_g(x)=\pi_g(x)-\pi_{g,\mathrm{tar}}(x)$, differentiating preference
scores~\eqref{eq:preference_scores}--\eqref{eq:bias_strength} gives
\begin{equation}
\begin{aligned}
\frac{\partial\widetilde{\mathcal{F}}}{\partial s_z}
&=\beta\bigl[\delta_F-\xi\,\delta_S-\delta_J\bigr],\quad
\frac{\partial\widetilde{\mathcal{F}}}{\partial s_x}
=\beta\,\delta_S,\\
\frac{\partial\widetilde{\mathcal{F}}}{\partial s_y}
&=\beta\bigl[-\delta_F+\xi\,\delta_S+\delta_J\bigr].
\end{aligned}
\label{eq:app_spin_grad}
\end{equation}
Adding the discrete Laplacian contribution
$2\lambda_{\mathrm{smooth}}(-\partial_x^2\mathbf{s})$ from the exchange term, and the
conservation-mapped term
$\lambda_{\mathrm{phys}}\partial_q R_{\mathrm{cons}}\cdot \pi_g(q_g-q)$ with
$R_{\mathrm{cons}}=(\partial_t\rho+\partial_x q)^2$ from the LWR residual,
recovers the update used in our solver.

\section{Hyperparameter Sensitivity Analysis}
\label{app:sensitivity}

\begin{figure*}[t]
\centering
\includegraphics[width=\textwidth]{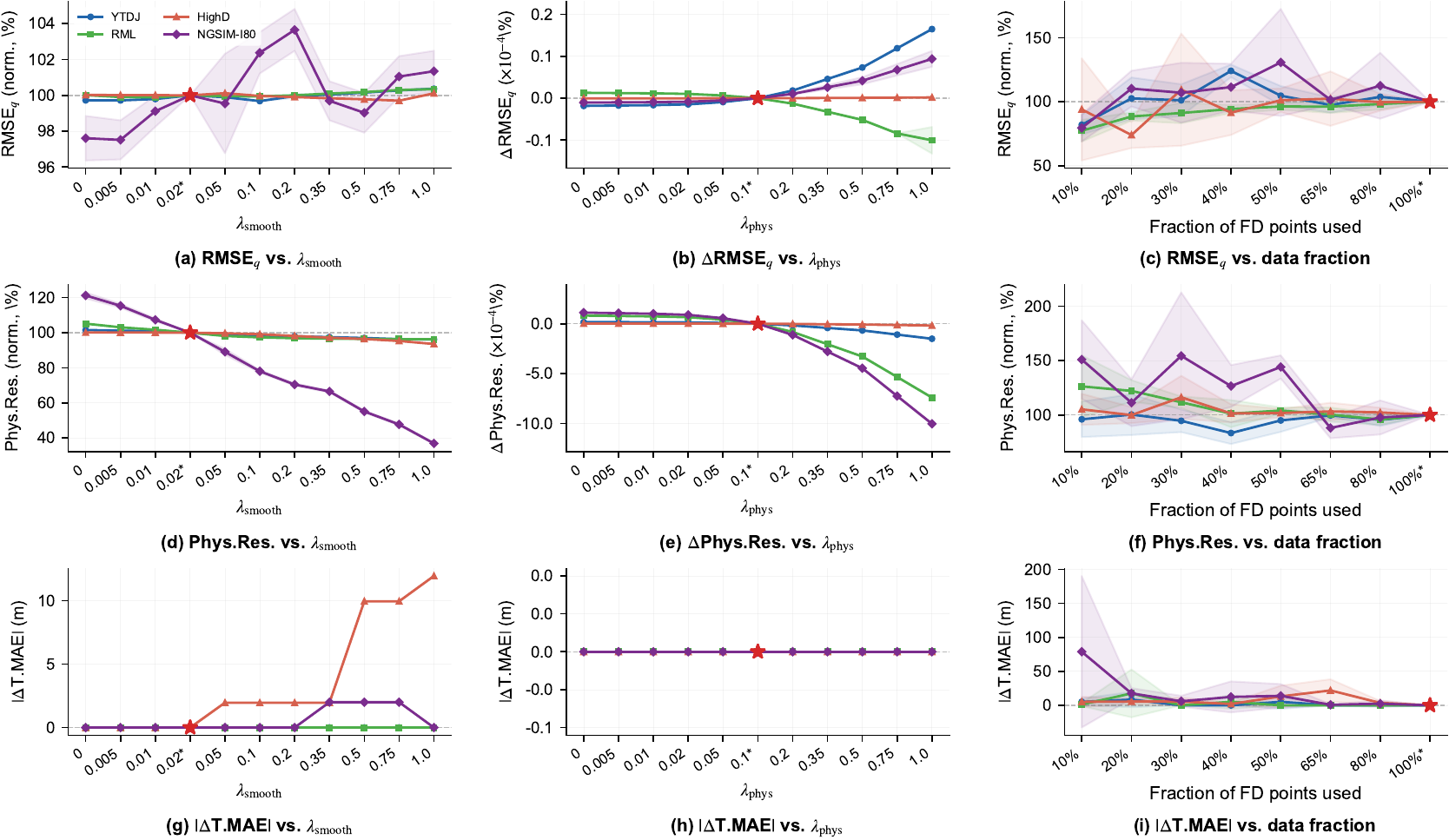}
\caption{Hyperparameter sensitivity to $\lambda_{\mathrm{smooth}}$, $\lambda_{\mathrm{phys}}$, and FD-point fraction. Red stars mark panel-specific defaults. In the $\lambda_{\mathrm{phys}}$ sweep, $\Delta$RMSE$_q$ and $\Delta$Phys.\,Res. are scaled as $\times10^{-4}\%$ to make near-flat variation visible.}
\label{fig:app_sensitivity}
\end{figure*}

Fig.~\ref{fig:app_sensitivity} isolates three perturbation axes around the Sec.~\ref{subsec:setup} defaults: $\lambda_{\mathrm{smooth}}\!\in\![0,1]$, $\lambda_{\mathrm{phys}}\!\in\![0,1]$, and FD-point fraction in $\{10\%,\ldots,100\%\}$. Rows report normalized $RMSE_q$, normalized Phys.\,Res., and $|\Delta\mathrm{T.MAE}|$ in meters, each measured relative to the panel-specific default.

The sweeps show three distinct behaviors. Increasing $\lambda_{\mathrm{smooth}}$ trades local fit for field regularity: YTDJ, RML, and HighD stay within a tight $RMSE_q$ band of $\pm$0.4\%, while NGSIM I-80 shows the largest and non-monotone response at about 4\%. At the same time, Phys.\,Res.\ drops as $\lambda_{\mathrm{smooth}}$ grows and reaches about 37\% of the default level on NGSIM I-80 at $\lambda_{\mathrm{smooth}}{=}1$, which is consistent with smoothing suppressing non-physical oscillations. By contrast, varying $\lambda_{\mathrm{phys}}$ leaves all three rows nearly unchanged, suggesting that near this operating point the conservation term acts as a weak E-step regularizer and FD-matching gradients dominate the update direction. Reducing FD-point fraction gives the strongest instability signal: on NGSIM I-80, $RMSE_q$ rises to about 131\% at 50\% coverage, and Phys.\,Res.\ rises to about 150\% on NGSIM I-80 and 126\% on RML at 10\% coverage. Overall, the red-star defaults lie in stable interior regions rather than boundary-sensitive edges.

\section{EM Convergence Analysis}
\label{app:convergence}

\begin{figure}[t]
\centering
\includegraphics[width=\columnwidth]{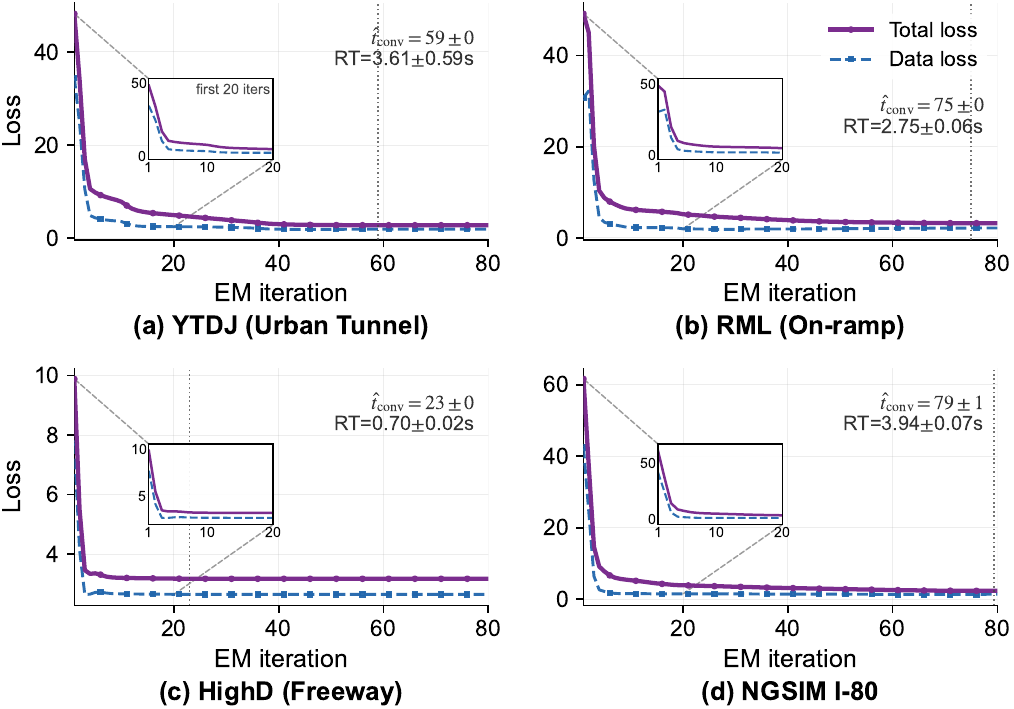}
\caption{EM convergence across four scenarios and five random seeds (2$\times$2 panel, single-column). Vertical marker: $\hat{t}_{\mathrm{conv}}$ (mean$\pm$std across seeds). RT denotes runtime in seconds (mean$\pm$std across seeds), reported from Table~\ref{tab:all_perf}.}
\label{fig:app_convergence}
\end{figure}

Fig.~\ref{fig:app_convergence} reports per-iteration augmented free-energy trajectories $\widetilde{\mathcal{F}}^{(t)}$ for all four scenarios across five evaluation seeds. Each panel shows seed-mean curves, with a solid line for $\widetilde{\mathcal{F}}^{(t)}$ and a dashed line for the data-fitting loss $\mathcal{L}_{q}^{(t)}$. The sharp early decline is driven by the data-fit term, after which smooth and physics regularization shape the flattening tail.

Convergence occurs within $\hat{t}_\mathrm{conv}=59\!\pm\!0$ iterations on YTDJ, $75\!\pm\!0$ on RML, $23\!\pm\!0$ on HighD, and $79\!\pm\!1$ on NGSIM I-80, consistent with the 23--79 range in Sec.~\ref{subsec:main_results}. The total augmented free-energy reduction at termination is $94.2\%$ on YTDJ, $93.4\%$ on RML, $67.9\%$ on HighD, and $96.1\%$ on NGSIM I-80. The smaller reduction on HighD comes from sparse supervision with $N{=}161$ FD points and a narrower initial energy range, rather than slower descent; the ratio $\mathcal{L}_q^{(\hat t_\mathrm{conv})}/\mathcal{L}_q^{(0)}$ remains below $0.4$. Cross-seed variability of $\hat{t}_\mathrm{conv}$ is at most one iteration on all datasets, indicating that once Phase Unfolding is established, the augmented free-energy landscape is effectively seed-insensitive.

\section{Baseline Implementation Details}
\label{app:baselines}

\medskip
\noindent\textbf{PWA-CTM}\quad
Each cell is hard-assigned to one of three density zones via the 33rd/67th percentiles of $\{\rho_{\mathrm{obs}}(x_i):\rho_{\mathrm{obs}}(x_i)>0\}$. Per-zone triangular FDs are calibrated by weighted least squares on the FD points that fall in that zone (avoiding the mismatch that arises when SpinFlow's soft prototypes are hard-assigned). At zone boundaries, the cell-wise indicator is replaced by an adaptive sigmoid blend
\[
\sigma_c=1/\!\bigl(1+e^{-4(c-c_b)/w_b}\bigr),
\]
where $c$ is the cell index, $c_b$ is the boundary-cell index between adjacent phases, and $w_b\in[2\%,12\%]$ of the total cell count is the adaptive blend width scaled by local density std (wider where transitions are uncertain). A stochastic metastability rule lets a boundary cell inherit its upstream neighbor with probability $0.15$.

\medskip
\noindent\textbf{VBGMM+KDE}\quad
A 3-component variational Bayesian Gaussian mixture (\texttt{sklearn} \texttt{BayesianGaussianMixture}, full covariance, \texttt{max\_iter}=500, $n_\mathrm{init}{=}3$) is fitted on normalized features $(\rho/\rho_{\max},q/q_{\max},x/x_{\max})$. Cluster--phase alignment is by ascending mean density. Posterior responsibilities are spatially smoothed by Gaussian kernel density estimation (KDE) with bandwidth $2\Delta x$ (evaluated at all $N\!\times\!M$ observation--cell pairs), then post-projected to the nearest prototype with weight $0.70\,\pi_{\mathrm{KDE}}+0.30\,\mathbf{1}_{\mathrm{hard}}$. Per-cluster triangular FDs are recalibrated from $\arg\max$ assignments.

\medskip
\noindent\textbf{PI-DeepONet}\quad
Branch network: $\mathbb{R}^2\!\to\!\mathbb{R}^{64}$, a three-layer Tanh multi-layer perceptron (MLP) with width 64 and input $(\rho/\rho_{\max},x/x_{\max})$. Trunk network: $\mathbb{R}^1\!\to\!\mathbb{R}^{64}$, same architecture with input $x/x_{\max}$. Output $\hat{q}=\mathbf{b}(\rho,x)\!\cdot\!\mathbf{t}(x)$. Training loss:
\[
\mathcal{L}_\mathrm{DON}=\tfrac{1}{N}\textstyle\sum_p(\hat{q}_p-q_p)^2+\lambda_{\mathrm{PDE}}\,\overline{(\partial_x q)^2},
\]
with $\lambda_{\mathrm{PDE}}=5.0$, Adam at learning rate $10^{-3}$, up to 2000 epochs, and rolling-window plateau detection using a window of 100 and a tolerance of $5\!\times\!10^{-3}$. The model outputs only $\hat q$, with no $\pi_g(x)$ or PED.

\section{Baseline Computational Complexity}
\label{app:complexity}

Let $N$ be the number of FD observation points, $M$ the spatial grid cells,
$K{=}20$ inner gradient steps per SpinFlow EM iteration,
$W_{\mathrm{net}}{=}64$ PI-DeepONet MLP hidden width, and $E$ the iteration or epoch budget.
Using the implementations in Appendix~\ref{app:baselines}, the total asymptotic costs are:

\medskip
\noindent\textbf{PWA-CTM}\quad
Phase assignment is deterministic and closed-form. Density-zone boundaries are fixed
by percentiles, each zone's triangular FD is solved once by WLS, and the sigmoid blend
is applied in one forward sweep over cells.
\[
  \mathcal{O}(N + M).
\]

\noindent\textbf{VBGMM+KDE}\quad
Variational EM runs for up to $E_\mathrm{gmm}{\le}500$ iterations at $\mathcal{O}(N)$
each with fixed $G{=}3$. The spatial KDE then evaluates a Gaussian kernel at all
$N{\times}M$ observation--cell pairs, which dominates in practice because
$NM\sim10^5 \gg E_\mathrm{gmm}N\sim10^4$.
\[
  \mathcal{O}(E_\mathrm{gmm}\,N + NM).
\]

\noindent\textbf{SpinFlow}\quad
Each EM iteration has an $\mathcal{O}(NM)$ kernel-aggregation term, which accumulates
a weighted sum over all $N$ observations for each of $M$ cells, plus an
$\mathcal{O}(KM)$ term from the $K$ inner spin-field gradient steps. Since
$K{=}20\!\ll\!N$, the per-iteration cost remains $\mathcal{O}(NM)$.
\[
  \mathcal{O}(\hat{t}_\mathrm{conv}\cdot NM).
\]

\noindent\textbf{PI-DeepONet}\quad
Each training epoch performs a forward--backward pass over $N{+}M$ points through
a three-layer MLP. Each dense layer contributes a $W_{\mathrm{net}}{\times}W_{\mathrm{net}}$
matrix product, and PDE autograd doubles the cost.
With $W_{\mathrm{net}}{=}64$, so $W_{\mathrm{net}}^2{=}4096$, and $N{+}M{\approx}700$, running to $E_\mathrm{don}{=}2000$
epochs yields roughly $5.7{\times}10^9$ scalar multiply-accumulate operations---a factor
absent from all other methods.
\[
  \mathcal{O}(E_\mathrm{don}\cdot(N+M)\cdot W_{\mathrm{net}}^2).
\]

\medskip
The derived orders predict the empirical runtime ordering
$\text{PWA-CTM}\!\ll\!\text{VBGMM+KDE}\!\lesssim\!\text{SpinFlow}\!\ll\!\text{PI-DeepONet}$
observed in Table~\ref{tab:all_perf}, spanning roughly two orders of magnitude from
${\sim}0.05$\,s to ${\sim}6$\,s.
Here RT refers to the runtime column in Table~\ref{tab:all_perf}; the larger values (e.g., 1420--2000) appear in the iteration column.
SpinFlow and VBGMM+KDE share the same $\mathcal{O}(NM)$ per-step kernel cost,
but SpinFlow repeats it for $\hat{t}_\mathrm{conv}{=}23\text{--}79$ physics-regularised
iterations, trading a modest runtime increase for calibrated phase weights
$\boldsymbol{\pi}(x)$ that neither VBGMM+KDE nor PI-DeepONet expose.


\end{document}